\documentclass[twocolumn]{aastex631}

\usepackage{booktabs}
\usepackage{amsmath}
\newcommand{\angstrom}{\mbox{\normalfont\AA}}
\renewcommand{\arraystretch}{1.15}
\shorttitle{Weak-CN Stars Are Ordinary RSGs}
\shortauthors{Ting et al.}

\begin{document}

\title{Weak-CN Stars Are Ordinary Cool Red Supergiants}

\author{Yuan-Sen Ting}
\affiliation{Department of Astronomy, The Ohio State University, Columbus, OH 43210, USA}
\affiliation{Center for Cosmology and AstroParticle Physics (CCAPP), The Ohio State University, Columbus, OH 43210, USA}
\affiliation{Max-Planck-Institut f\"ur Astronomie, K\"onigstuhl 17, D-69117 Heidelberg, Germany}

\author{Puragra Guhathakurta}
\affiliation{Department of Astronomy and Astrophysics, University of California Santa Cruz, 1156 High Street, Santa Cruz, CA 95064, USA}

\author{Douglas Grion Filho}
\affiliation{Department of Astronomy and Astrophysics, University of California Santa Cruz, 1156 High Street, Santa Cruz, CA 95064, USA}

\author{Evan N.\ Kirby}
\affiliation{Department of Physics and Astronomy, University of Notre Dame, 225 Nieuwland Science Hall, Notre Dame, IN 46556, USA}

\author{Elliot M.\ Kim}
\affiliation{Department of Computer Science, Cornell University, Ithaca, NY 14853, USA}

\begin{abstract}
Weak CN absorption near $\sim 8000\,$\angstrom\ has recently been detected in evolved red supergiants (RSGs) of $5$--$10\,\mathrm{M}_\odot$ across three Local Group galaxies \citep{guh25,gri25}.  These weak-CN RSGs sit in a narrow molecular regime: cool enough for CN to be visible in a non-carbon, C/O$<1$ atmosphere, but warm enough that TiO is not saturated and changes in $T_{\rm eff}$ and in the surface C+N reservoir move CN and TiO in distinct directions.  We test this picture with pseudo-continuum equivalent widths (EWs) measured from LMC, M33, and M31 weak-CN and carbon-star coadds, compared at matched resolution to a self-consistent grid of synthetic RSG atmospheres spanning $T_{\rm eff}$, $[\alpha/{\rm Fe}]$, and surface C and N offsets relative to each host's scaled-solar baseline.  Ordinary cool-RSG models reproduce the weak-CN coadds across all three hosts, with per-feature residuals at the level of the adopted EW systematic floors.  The robust observable is the combined surface abundance $\Delta\mathrm{[C/H]}+\Delta\mathrm{[N/H]}$ rather than each offset individually, because CN forms from the product of available C and N number densities.  Mapping $\Delta\mathrm{[C/H]}+\Delta\mathrm{[N/H]}$ to initial rotation through PARSEC v2.0 has modest leverage --- the variable shifts by $\simeq 0.07\,$dex from $\omega_i=0$ to $\omega_i=0.6$ --- and within this resolution slow-rotation first dredge-up is consistent with LMC and M33, and with M31 once a single-feature CaT\,8542\,\angstrom\ calibration anchor is allowed.  The straightforward resolution of the discovery puzzle is therefore that weak CN is not an exotic carbon-star intermediate but the expected molecular-equilibrium signature of ordinary cool RSGs.
\end{abstract}


\section{Introduction}\label{sec:intro}

RSGs dominate the integrated optical and near-infrared light of young stellar populations during the first $\sim 50\,$Myr after a star-formation event, and they are the principal sources returning CNO-cycle-processed material to the interstellar medium on that timescale.  Their surface abundances are therefore a direct input to stellar-population synthesis and Galactic chemical evolution.  Measuring those abundances at high resolution is difficult: cool-RSG optical spectra are blanketed by a dense molecular forest \citep{lev05,gus08}, the photospheres pulsate and convect on month timescales, and the apertures required at high signal-to-noise reach only the brightest nearest stars.  An indirect spectroscopic tracer that responds robustly to RSG CNO chemistry at low or moderate spectral resolution would be valuable, and it is in this context that we re-examine the recently identified weak-CN class.

The weak-CN stars discovered by \citet{guh25} and extended to the LMC by \citet{gri25} occupy an awkward spectroscopic category.  Their spectra show a shallow CN absorption band around $7800$--$8200\,\angstrom$, the same broad feature commonly used to identify carbon stars \citep{boyer13,hamren16,past20}, but the rest of the optical spectrum has the morphology of a non-carbon, C/O$<1$ cool star, with prominent TiO bands.  The stars sit on the young red-supergiant locus of the color--magnitude diagram, at masses $5$--$10\,\mathrm{M}_\odot$ and main-sequence ages $\simeq 40\,$Myr in the core-He-burning phase.  \citet{guh25} treated the class as mysterious: CN was the carbon-star-like feature, while TiO was the normal non-carbon-star feature, sitting on the cool-RSG locus, whose effective-temperature scale in this regime was established by \citet{lev05}.  \citet{gri25} showed that similar objects are common in the LMC, used the pre-computed \citet{ari16} model atmospheres to bracket $T_{\rm eff}\simeq 4000$--$4600\,$K, and argued that weak CN can occur in normal evolved stars over a restricted temperature and gravity range, with surface nitrogen enhancement from main-sequence rotational mixing as a plausible source of EW scatter.

The cool-RSG optical spectrum is dominated by two molecules, TiO and CN, whose responses to temperature and chemistry are sharply different.  TiO crosses its dissociation transition near $T_{\rm eff}\sim 3800\,$K, well inside the WCN plateau, while CN remains effective up to $T_{\rm eff}\sim 4300\,$K.  CN is also a C--N molecule whose strength depends directly on the surface C and N reservoirs, whereas TiO is essentially insensitive to those abundances at fixed metallicity.  This asymmetry between the two molecules is the physical basis on which the rest of the analysis rests.

Two mechanisms can therefore produce a shallow CN feature in such a spectrum, and the discovery-paper ambiguity reflects exactly that.  The first is a $T_{\rm eff}$ selection: in the temperature range where TiO weakens against the molecular pseudo-continuum, the residual CN band becomes a natural feature of an ordinary C/O$<$1 atmosphere, with no abundance peculiarity required.  The second is a surface CNO enhancement from first dredge-up plus main-sequence rotational mixing, which raises the surface nitrogen abundance and deepens CN at fixed $T_{\rm eff}$.  The CN band alone cannot distinguish these two scenarios; both produce a shallow CN morphology of comparable strength.

The diagnostic leverage comes from the partnership between CN and TiO: $T_{\rm eff}$ moves TiO strongly but barely touches CN, whereas a CNO surface enhancement moves CN through the C and N reservoirs but barely shifts TiO at fixed $T_{\rm eff}$.  A joint $\chi^2$ over the two indices therefore separates temperature from chemistry.  Two further pieces of physics shape how that separation works in detail: the photospheric structure back-reacts to abundance perturbations through molecular line blanketing at $\log g\simeq 0$ (Section~\ref{sec:grid}), and the CN bands respond to the product of the C and N reservoirs so the constrained surface variable is the combination $\Delta\mathrm{[C/H]}+\Delta\mathrm{[N/H]}$ rather than each offset individually (Section~\ref{sec:physics}).  Mapping that combination onto initial rotation rate further requires a stellar-evolution prior.

The paper is organised as follows.  Section~\ref{sec:data} describes the observed coadds and the resolution-consistent EW measurement.  Section~\ref{sec:grid} describes the self-consistent atmosphere grid.  Section~\ref{sec:physics} uses the model atmospheres to show why weak CN is a normal cool-RSG feature.  Section~\ref{sec:results} presents the per-galaxy joint $\chi^2$ fits, the best-fit spectra, and the per-galaxy inferred surface composition.  Section~\ref{sec:discussion} discusses the C+N degeneracy, where in the EW plane the diagnostic discriminates rotation, the role of C$_2$ Swan as a candidate pure-carbon lever, the slow-rotation consistency of the data, and the implications for using weak-CN stars as recent-SFR tracers.

\section{Data and EW Measurement}\label{sec:data}

\begin{figure*}[!htbp]
\centering
\includegraphics[width=\textwidth]{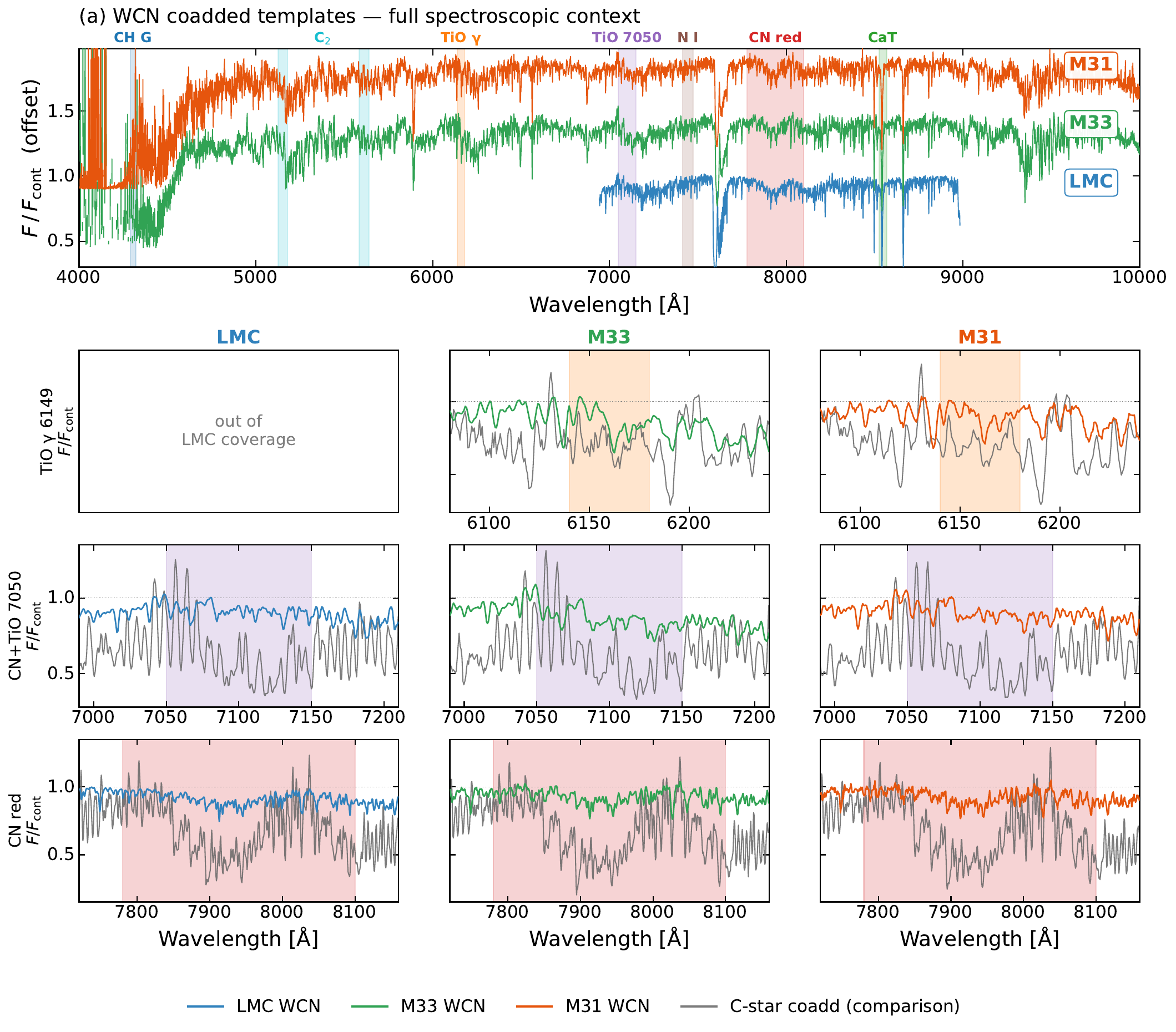}
\caption{Observed weak-CN and carbon-star coadded templates.  The top panel shows the three weak-CN coadds on a common pseudo-continuum scale; the $3\times 3$ zooms compare each galaxy's weak-CN coadd (coloured) with its carbon-star coadd (grey) in the three diagnostic windows TiO $\gamma$ 6149\,\angstrom, CN+TiO 7050\,\angstrom, and CN red.  The weak-CN class is what the discovery papers noticed: a CN band that has the carbon-star morphology but only $\sim 20$--$30\%$ of its depth, while the surrounding TiO indices retain the strength expected of a normal C/O$<$1 cool atmosphere.  This is the spectroscopic ambiguity that the rest of the paper resolves.}
\label{fig:obs}
\end{figure*}

We use the inverse-variance-weighted coadded templates of the visually classified weak-CN and carbon-star samples published by \citet{guh25} for M31 and M33 and by \citet{gri25} for the LMC.  In both papers the weak-CN and carbon-star populations are drawn from the same spectroscopic surveys (Keck/DEIMOS for M31 and M33, CTIO/Hydra for the LMC) and the carbon-star coadd is the classical comparison template against which the weak-CN class was originally defined --- a deeper CN red band and a clear C$_2$ Swan absorption.  We adopt those coadds as inputs without re-classification.  The high-quality subsets used here draw from 75 weak-CN and 71 carbon-star spectra in M31, 140 and 34 in M33, and 779 and 1950 in the LMC, a cleaner subset of the broader survey-level identifications of $224 / 659 / 779$ weak-CN stars in M31\,/\,M33\,/\,LMC.  M31 and M33 spectra were taken with Keck/DEIMOS at native resolution $R\simeq 2000$ over $4000$--$10\,000\,\angstrom$; the LMC spectra were taken with CTIO/Hydra at $R\simeq 7000$ over $6850$--$9150\,\angstrom$.  Figure~\ref{fig:obs} shows the observed templates and the EW windows used here.

The purpose of the carbon-star overlay is historical and diagnostic: the weak-CN class was noticed because its broad CN morphology resembles a diluted carbon-star CN band, but the same spectra retain non-carbon-star features such as TiO and CaT.  The carbon-star coadds therefore define what the weak-CN stars are not.

The figure also shows a practical limitation that carries through the analysis.  The LMC HYDRA coverage contains TiO 7050\,\angstrom, CN red, N\,{\sc i}, and CaT, but does not extend to the bluer TiO\,$\gamma$ or C$_2$ Swan windows that M31 and M33 do (at lower spectral resolution).  We therefore do not force all three galaxies into an identical feature set; instead we use the features each data set can support and keep the interpretation tied to the available diagnostic geometry.

All EWs are measured after division by the same pseudo-continuum recipe in data and models.  For each spectrum we estimate a local continuum $F_c(\lambda)$ as the 95th percentile flux in a rolling $200\,\angstrom$ window, followed by a light Gaussian smoothing of that upper envelope.  The broad window is intentional: in cool RSGs there is no true line-free continuum in the optical molecular forest \citep{gus08}, so the observable must be a pseudo-continuum index rather than a classical isolated-line EW.  The typical RSG molecular bandhead spacing of several tens of \angstrom\ requires the wider window adopted here.  For a feature window $[\lambda_1,\lambda_2]$ we then measure
\begin{equation}
{\rm EW} =
\int_{\lambda_1}^{\lambda_2}
\left[1 - \frac{F(\lambda)}{F_c(\lambda)}\right]\,d\lambda ,
\label{eq:ew}
\end{equation}
using the finite pixels in the window.  Positive EW therefore corresponds to absorption below the pseudo-continuum.

The equally important technical point is resolution consistency.  The synthetic spectra are computed at high resolution, where inter-line gaps in the molecular forest remain visible; the observations are at lower resolution, where those gaps are blended into the band profiles.  Because the continuum estimate in Equation~\ref{eq:ew} is nonlinear, convolution and normalization do not commute.  We therefore convolve every model spectrum to the relevant observational resolution before applying the continuum and EW measurement: $R\simeq 7000$ for the LMC/HYDRA spectra and $R\simeq 2000$ for the M31/M33 DEIMOS spectra.  The observed spectra are processed at their native resolution.

Each model spectrum is therefore processed through the same sequence as the data: the available synthetic wavelength pieces are merged, the merged spectrum is convolved from the native synthesis resolution to the observation's resolution, and only then is the pseudo-continuum estimated and the EW integrated.  The LMC HYDRA spectra are kept at their native $R\simeq 7000$ so that model and data are compared at matched effective resolution.

The wavelength windows and adopted uncertainty floors are listed in Table~\ref{tab:index}.  The sideband definitions used for bookkeeping are not used as a local straight-line continuum in the final EW; the pseudo-continuum in Equation~\ref{eq:ew} is measured from the full local spectrum.  We keep the conventional name ``TiO 7050'' for the 7050--7150\,\angstrom\ index, but in these cool-RSG spectra it should be read as a CN+TiO blend rather than a pure TiO band.  Counting transitions in the CN A--X red-system line list of \citet{brooke14} shows that the $7000$--$7200\,$\angstrom\ window contains $\sim 21{,}000$ CN lines with $\sum gf \approx 193$, while the canonical CN red window at $7780$--$8100\,$\angstrom\ contains $\sim 24{,}000$ lines with $\sum gf \approx 259$.  The CN contribution to ``TiO 7050'' is therefore at $\sim 75\%$ of the canonical CN-red level, so the index is sensitive to both molecular species and to abundance changes in the C+N reservoir.

\begin{table*}[!htbp]
\centering
\caption{EW index definitions and fit uncertainty floors.  $\sigma_{\rm EW}$ values are constructed in quadrature from a per-(galaxy, feature) bootstrap of the pseudo-continuum recipe and a 20\% line-list calibration term (Equation~\ref{eq:sigma}).  TiO 7050\,\angstrom\ is a CN$+$TiO blend so it is used in the $\chi^2$ fit only for the LMC (where TiO $\gamma$ is out of HYDRA coverage); for M31 and M33 the cleaner TiO $\gamma$ anchor is used in the fit while TiO 7050\,\angstrom\ is retained as a diagnostic comparison in Figure~\ref{fig:locus}.}
\label{tab:index}
\small
\begin{tabular}{llcccl}
\toprule
Index & Window (\angstrom) & \multicolumn{3}{c}{$\sigma_{\rm EW}$ (\angstrom)} & Used for \\
      &                    & LMC & M33 & M31 & \\
\midrule
C$_2$ Swan 5165\,\angstrom & 5125--5180 & \nodata & 3.4 & 3.4 & M31, M33 \\
C$_2$ Swan 5636\,\angstrom & 5585--5640 & \nodata & 2.6 & 2.5 & M31, M33 \\
TiO $\gamma$ 6149\,\angstrom & 6140--6180 & \nodata & 1.2 & 0.9 & M31, M33 \\
TiO 7050\,\angstrom (CN+TiO blend) & 7050--7150 & 2.6 & 3.0 & 2.1 & LMC fit only (Fig.~\ref{fig:locus} for M31, M33) \\
N~{\sc i} 7442\,\angstrom  & 7415--7475 & 0.7 & 1.0 & 0.9 & all \\
CN red          & 7780--8100 & 7.4 & 4.5 & 4.8 & all \\
CaT 8542\,\angstrom        & 8525--8568 & 1.6 & 1.6 & 1.5 & all \\
\bottomrule
\end{tabular}
\end{table*}

The $\sigma_{\rm EW}$ values in Table~\ref{tab:index} are systematic floors rather than formal photon-noise errors --- the high-quality coadds aggregate of order $10^2$ stars per galaxy and have negligible photon noise per pixel at the coadd level.  We construct them in quadrature from two terms, each measured directly:
\begin{equation}
\sigma_{\rm EW}^2 \;=\; \sigma_{\rm recipe}^2 \;+\; \sigma_{\rm LL}^2 ,
\label{eq:sigma}
\end{equation}
where $\sigma_{\rm recipe}$ is the residual systematic from the pseudo-continuum recipe and $\sigma_{\rm LL}$ is a line-list calibration term.

For $\sigma_{\rm recipe}$ we bootstrap the recipe over a $4\times 6$ grid of window widths (150--300\,\angstrom) and percentiles (92--98\%), apply each combination to both the convolved model and the observation, and take the per-feature standard deviation of the resulting (EW$_{\rm model}-$EW$_{\rm obs}$) distribution.  This isolates the recipe drift that does not cancel in the model-vs-data comparison.  $\sigma_{\rm recipe}$ is largest for the LMC HYDRA coadd (4.7\,\angstrom\ for CN red), where the higher native resolution ($R\simeq 7000$) leaves more line structure for the rolling-percentile filter to land on, and falls to $0.1$--$1.7\,\angstrom$ across all features for the lower-resolution M31/M33 DEIMOS coadds.

{\sloppy For $\sigma_{\rm LL}$ we adopt a 20\% relative line-list calibration term, $\sigma_{\rm LL}=0.2\,|\mathrm{EW}_{\rm obs}|$.  The CN A--X red-system list of \citet{brooke14}, the Schwenke TiO list, and the C$_2$ Swan opacity are calibrated primarily on Galactic stellar samples; the 20\% fraction encompasses oscillator-strength uncertainty, NLTE departures for atomic features (the N\,{\sc i} and Ca\,{\sc ii} lines in particular), and local-blending residuals at the host metallicities here.\footnote{We have repeated the joint fit with $\sigma_{\rm LL}$ set to 10\%, 15\%, 25\%, and 30\%, and with separate fractions for atomic and molecular features.  The inferred $\Delta\mathrm{C}+\Delta\mathrm{N}$ shifts by at most $0.2\,$dex in any galaxy and the location of the $\chi^2$ minimum is essentially unchanged.  The \emph{absolute} value of the joint EW $\chi^2$, however, is not similarly invariant --- $\sigma_{\rm LL}$ enters the denominator, so smaller $\sigma_{\rm LL}$ inflates $\chi^2/N_{\rm feat}$ proportionally.  We therefore use the $\chi^2$ surface to locate and rank grid points (where it is approximately $\sigma_{\rm LL}$-invariant) and read the absolute goodness-of-fit from the per-feature $\sigma$ residuals (Section~\ref{sec:results-fits}) rather than from $\chi^2/N_{\rm feat}$ itself.}  Combining the two terms gives the per-galaxy, per-feature floors of Table~\ref{tab:index}.  $\sigma_{\rm EW}$ is dominated by $\sigma_{\rm recipe}$ for the LMC's high-resolution CN red and TiO 7050 indices, and by $\sigma_{\rm LL}$ everywhere else.\par}

The measured EWs are given in Table~\ref{tab:obs}.  The principal contrast that defines the weak-CN class is the CN red index: the carbon-star coadds give $\simeq 102$--$104\,\angstrom$ across all three galaxies, while the weak-CN coadds give $22.6$--$28.7\,\angstrom$, a factor of three to four shallower.  The TiO and Ca\,{\sc ii} indices show the opposite behaviour expected of C/O$<$1 atmospheres relative to C-stars; the TiO 7050\,\angstrom\ EW in the carbon-star coadds is also large at $\simeq 38$--$40\,\angstrom$, but in that window the carbon-star value is dominated by the blue wing of the C$_2$ Swan band rather than by genuine TiO absorption.

\begin{table*}[!htbp]
\centering
\caption{Observed equivalent widths from the current measurement pipeline.  Per-(galaxy, feature) $\sigma_{\rm EW}$ floors used in the fits are given in Table~\ref{tab:index}.  Blank entries are outside the wavelength coverage or not used in the fit.}
\label{tab:obs}
\small
\begin{tabular}{llrrrrrrr}
\toprule
Galaxy & Class & C$_2$ 5165\,\AA & C$_2$ 5636\,\AA & TiO $\gamma$ & TiO 7050\,\AA & N~{\sc i} & CN red & CaT \\
       &       & \multicolumn{7}{c}{(\AA)} \\
\midrule
LMC & WCN & \nodata & \nodata & \nodata & 8.8 & 3.4 & 28.7 & 7.8 \\
LMC & C star & \nodata & \nodata & \nodata & 38.8 & 16.4 & 102.2 & 14.6 \\
M33 & WCN & 15.4 & 9.9 & 5.5 & 14.7 & 5.0 & 22.6 & 8.1 \\
M33 & C star & 42.4 & 31.8 & 11.2 & 38.1 & 16.1 & 103.7 & 15.0 \\
M31 & WCN & 14.6 & 9.3 & 4.0 & 10.5 & 4.5 & 24.0 & 7.4 \\
M31 & C star & 39.7 & 32.8 & 11.6 & 39.6 & 17.0 & 103.7 & 15.0 \\
\bottomrule
\end{tabular}
\end{table*}

\section{Model Grid}\label{sec:grid}

We compute self-consistent cool-RSG model atmospheres and synthetic spectra using \texttt{pykurucz} \citep{kim26}, a pure-Python re-implementation of the Kurucz ATLAS12 atmosphere iteration and SYNTHE line-synthesis pipeline \citep{kur05}.  The grid fixes $\log g=0$ and microturbulence at $2\,{\rm km\,s^{-1}}$, and spans host metallicity, $T_{\rm eff}$, $[\alpha/{\rm Fe}]$, and C/N surface offsets (Table~\ref{tab:grid}).  The main synthesis covers the red optical windows used for CN, TiO 7050\,\angstrom, N~{\sc i}, and CaT.  Additional M31/M33 synthesis extensions cover the C$_2$ Swan windows as C-sensitive sanity checks.

The relevant distinction is that this is an ATLAS12-style direct-opacity-sampling calculation rather than an ATLAS9 opacity-distribution-function interpolation.  Each atmosphere is iterated with the requested abundance pattern in place and with molecules enabled: the iteration solves Saha--Boltzmann populations and molecular equilibrium, continuous opacity, atomic and molecular line opacity, the Rosseland mean optical-depth scale, radiative transfer, mixing-length convection where unstable, and temperature corrections until a converged Kurucz-format atmosphere is produced.  The SYNTHE step then synthesizes the emergent spectrum on a high-resolution wavelength grid using the same classes of opacity data as the Kurucz pipeline, including GFALL atomic lines, Schwenke TiO, and the diatomic molecular lists relevant here (CN, CO, C$_2$, CH, OH, MgH, FeH, and others).  End-to-end validation against the original Fortran ATLAS12+SYNTHE pipeline gives median normalized-flux differences below $0.005\%$ and 99th-percentile differences below $0.3\%$ over representative test cases \citep{kim26}, well below the EW systematic floors adopted here.

The reason we do not use pre-computed scaled-solar grids \citep[such as][]{ari16} as the engine is that, at $\log g \simeq 0$, the molecular contribution to the Rosseland mean opacity at the formation depth is a substantial fraction of the total in MARCS calculations \citep{gus08}, compared to a few per cent at K-giant gravities of similar $T_{\rm eff}$.  Perturbing C, N, or $[\alpha/{\rm Fe}]$ at RSG gravities therefore changes the line opacity at depth substantially, which back-reacts on the temperature--pressure structure through line blanketing.  At K-giant gravities the molecular fraction is small enough that an abundance perturbation can be applied at fixed atmosphere structure --- this is the operating regime of optical CN-based abundance work such as \citet{tin18} --- but at RSG gravities the back-reaction can move the formation-layer temperature by tens of K and is not negligible.

\texttt{pykurucz} accepts arbitrary per-element abundance overrides and iterates the atmosphere to radiative--convective equilibrium under the perturbed opacity at every grid point, so the model T--P profile responds to the imposed CNO and $[\alpha/{\rm Fe}]$ offsets and the resulting EWs reflect that back-reaction rather than a fixed-structure approximation.

The C and N offsets are defined relative to each host's scaled-solar baseline, not relative to a common absolute abundance.  This convention matters for the interpretation of $\Delta\mathrm{C}+\Delta\mathrm{N}$: it is a combined C+N surface abundance offset at fixed host metallicity, not a statement that the three galaxies have the same initial CNO pattern.  The $[\alpha/{\rm Fe}]$ axis is included because oxygen-bearing opacity and TiO strength can change both the molecular equilibrium and the atmospheric temperature structure.  We use the full grid in the fit rather than imposing a temperature or $[\alpha/{\rm Fe}]$ prior from the outset.

\begin{table*}[!htbp]
\centering
\caption{Synthetic grid and fitted feature sets.  Host [M/H] values are from \citet{harris09,zar04} for the LMC, \citet{mar09} for M33, and \citet{dalc12} for M31.}
\label{tab:grid}
\small
\begin{tabular}{ll}
\toprule
Quantity & Values \\
\midrule
$T_{\rm eff}$ & 3800--4600 K, 100 K spacing \\
$\Delta{\rm [C/H]}$ & $-0.6, -0.5, -0.4, -0.2, -0.1, 0.0, +0.1, +0.2, +0.3$ \\
$\Delta{\rm [N/H]}$ & $0.0, +0.1, +0.2, +0.3, +0.5, +0.6, +0.7, +0.8$ \\
$\Delta[\alpha/{\rm Fe}]$ & -0.3, 0.0, +0.3 \\
Host [M/H] & LMC $-0.5$; M33 $-0.3$; M31 $\sim 0$ \\
Observation $R$ & LMC 7000; M33/M31 2000 \\
\midrule
LMC fit windows & CN red, TiO 7050, CaT, N~{\sc i} \\
M33/M31 fit windows & CN red, TiO $\gamma$, CaT, N~{\sc i}, C$_2$ 5636, C$_2$ 5165 \\
\bottomrule
\end{tabular}
\end{table*}

\section{Molecular Physics Context}\label{sec:physics}

\begin{figure*}[!htbp]
\centering
\includegraphics[width=\textwidth]{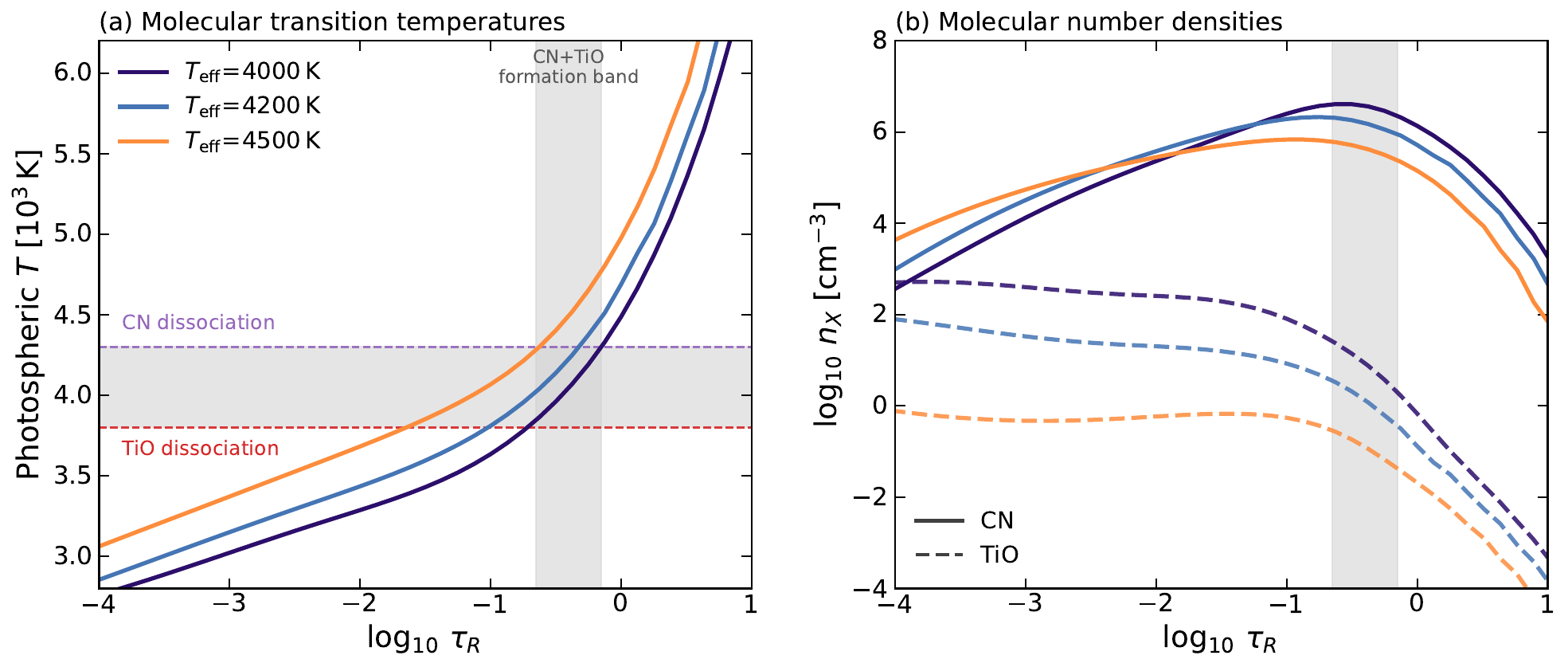}
\caption{Atmospheric physics underlying the weak-CN diagnostic.  Left: $T(\tau_R)$ for four representative cool-RSG models, with the TiO and CN dissociation transitions marked and a common Rosseland-depth reference band at $\log\tau_R \simeq -0.4$.  Right: CN and TiO number densities at the same models on the same depth axis.  The figure shows the orthogonal response of the two molecules: as $T_{\rm eff}$ drops through the WCN range, TiO grows steeply while CN grows gently, whereas a CNO surface enhancement acts directly on CN through the C and N reservoirs but barely shifts TiO at fixed $T_{\rm eff}$.  Temperature and surface chemistry therefore move the two indices in distinct directions, which is what allows the joint $\chi^2$ fit of Section~\ref{sec:results} to break the $T_{\rm eff}$--CNO degeneracy that a CN-only measurement cannot.}
\label{fig:atm}
\end{figure*}

With the self-consistent grid in hand, we now use it to understand why the joint EW fit can separate $T_{\rm eff}$ from surface CNO in the cool-RSG regime, and what surface variables the optical CN and TiO indices can and cannot constrain.  Figure~\ref{fig:atm} is the physical backdrop for that argument: the four representative cool-RSG models visualise the $T(\tau_R)$ structure and the CN and TiO number densities on a common Rosseland-depth axis centred at $\log_{10}\tau_R=-0.4$, with a half-width of $0.25\,$dex.  That band is where the optical molecular pseudo-continuum in these cool, low-gravity atmospheres emerges --- near the photosphere rather than at the outer boundary or in the deep diffusion regime --- so the T--P structure and the molecular number densities relevant to the observed EWs can be read together at the same depth.

The useful physical scale is set by the molecular dissociation temperatures.  TiO has bond dissociation energy $D_0\simeq 6.9\,{\rm eV}$ and crosses its formation-layer transition around $T\simeq 3800\,$K, while CN has $D_0\simeq 7.7\,{\rm eV}$ and crosses near $T\simeq 4300\,$K --- a $\sim 500\,$K offset that puts the two molecules out of phase across the cool-RSG plateau.  Across the formation band, a $4500\,$K model is relatively molecule-poor; by $4200\,$K CN has grown substantially and TiO is beginning to grow faster; by $4000\,$K TiO has grown disproportionately while CN is still present rather than erased.  A change in $T_{\rm eff}$ can therefore move the TiO indices strongly without erasing the red CN band, and an atmosphere can show weak but recognisable CN while keeping non-carbon-star RSG morphology --- which is the physical reason the C-star comparison in Figure~\ref{fig:obs} is useful for identifying the feature but misleading if interpreted as evidence for C-star chemistry.

The same molecular-equilibrium argument fixes what the CN band can and cannot tell us about the surface chemistry.  CN forms from the C and N populations that remain after CO sequesters most of the available carbon, and the column density of the molecule scales as $n_\mathrm{C}\cdot n_\mathrm{N}$ to first approximation.  A CN-only EW measurement at fixed $T_{\rm eff}$ therefore constrains the product $n_\mathrm{C}\cdot n_\mathrm{N}$ rather than $n_\mathrm{C}$ and $n_\mathrm{N}$ separately; in abundance space this is the combined surface offset $\Delta\mathrm{[C/H]}+\Delta\mathrm{[N/H]}$ along a degeneracy ridge in $(\Delta\mathrm{C}, \Delta\mathrm{N})$.

TiO does not break the degeneracy because it is essentially insensitive to the C and N reservoirs at fixed metallicity, and the optical C$_2$ Swan features, which would in principle anchor $\Delta\mathrm{C}$ independently, are at low-resolution sensitivity floors that limit their leverage (Section~\ref{sec:disc-c2}).  $\Delta\mathrm{C}+\Delta\mathrm{N}$ is therefore the surface variable the optical EW fit measures cleanly, and the individual $\Delta\mathrm{C}$ and $\Delta\mathrm{N}$ values returned by the formal $\chi^2$ argmin should be read as a representative point on a broader allowed locus rather than as unique abundance measurements.

\section{Results}\label{sec:results}

\subsection{Where the Diagnostic Discriminates}\label{sec:results-geometry}

\begin{figure*}[!htbp]
\centering
\includegraphics[width=\textwidth]{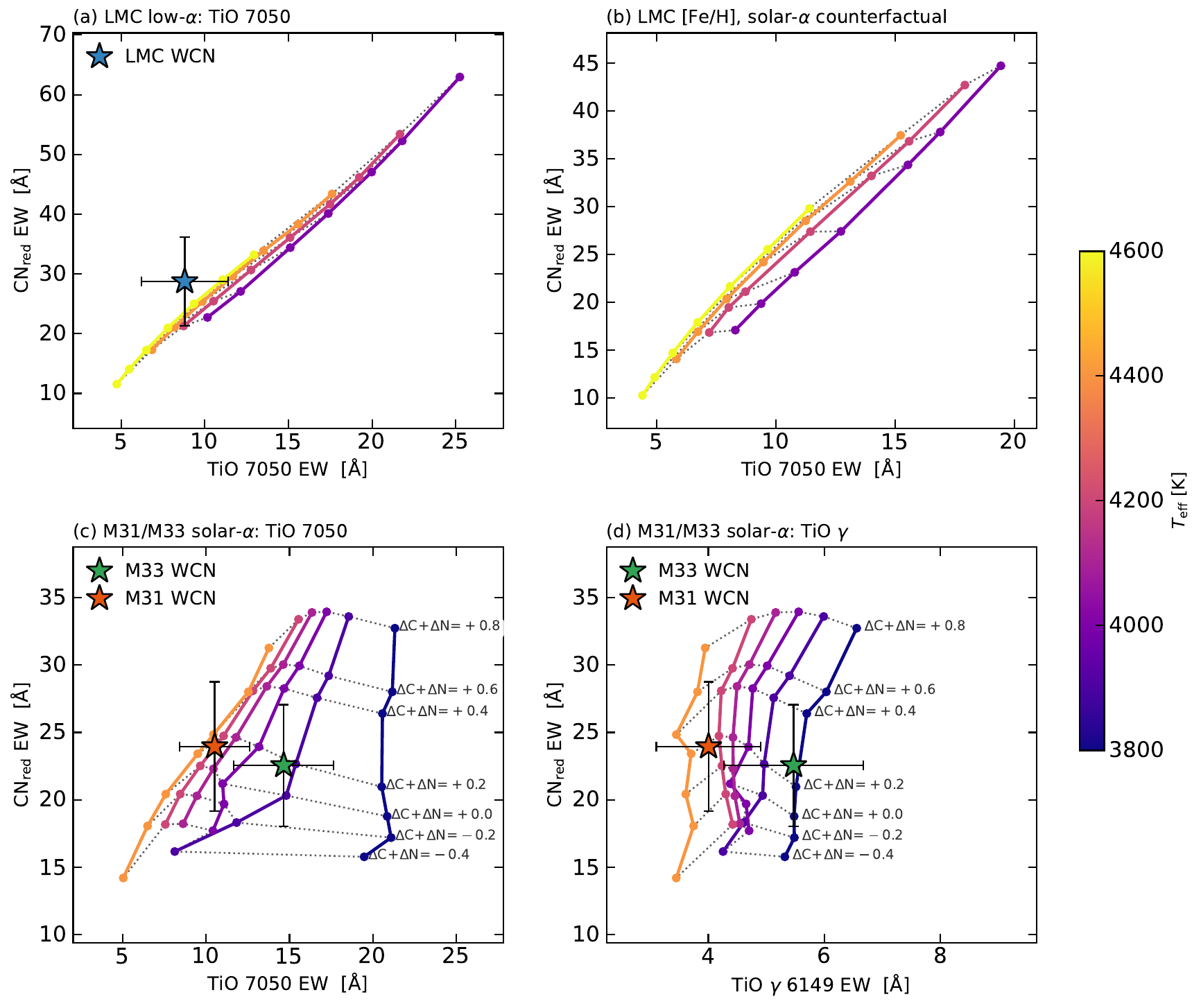}
\caption{Where the diagnostic discriminates parameters in the EW plane.  Symbols are the observed weak-CN coadds; coloured curves are model iso-$T_{\rm eff}$ tracks computed at each host's metallicity and $[\alpha/{\rm Fe}]$, with each plotted point obtained by averaging the model EWs over the grid points whose $\Delta\mathrm{[C/H]}+\Delta\mathrm{[N/H]}$ falls within $\pm 0.05\,$dex of the fixed sum.  Dotted curves connect fixed $\Delta\mathrm{[C/H]}+\Delta\mathrm{[N/H]}$ across $T_{\rm eff}$ by the same averaging recipe.  The bottom panels label the fixed sums, sampled at $-0.4,-0.2,0.0,+0.2,+0.4,+0.6$, and $+0.8\,$dex; the top panels omit those labels for readability.  In the M33 and M31 solar-$\alpha$ TiO 7050\,\angstrom\ plane the iso-$T_{\rm eff}$ and iso-$(\Delta\mathrm{C}+\Delta\mathrm{N})$ families fan out and moderate positive $\Delta\mathrm{C}+\Delta\mathrm{N}$ values overlap the weak-CN data; the TiO\,$\gamma$ plane provides a cleaner TiO consistency check with the same qualitative placement.  In the LMC low-$\alpha$ grid, lower [O/Fe] and lower [Fe/H] keep CN and TiO closer to the linear curve-of-growth and the families collapse, so the LMC weak-CN data point falls on the locus but the optical indices are a poor discriminator of the CNO-processing direction.}
\label{fig:locus}
\end{figure*}

Before running the formal joint $\chi^2$ fit, we examine where the observed weak-CN coadds fall in the model EW plane and how cleanly the data separate temperature from CNO chemistry there.  The orthogonality argument of Section~\ref{sec:physics} guarantees that this separation exists in principle; whether it is operative in any specific host depends on how the iso-$T_{\rm eff}$ and iso-$(\Delta\mathrm{C}+\Delta\mathrm{N})$ families project onto the observed (TiO 7050\,\angstrom, CN red) plane (Figure~\ref{fig:locus}).

In the M33 and M31 grids, both at near-solar $[\alpha/{\rm Fe}]$, the iso-$T_{\rm eff}$ and iso-$(\Delta\mathrm{C}+\Delta\mathrm{N})$ families fan out, and moderate positive values of $\Delta\mathrm{C}+\Delta\mathrm{N}$ ($\simeq 0$ to $+0.3$ dex) occupy the same part of the plane as the observed weak-CN coadds.  TiO\,$\gamma$, which is less affected by CN contamination than the 7050 blend (Section~\ref{sec:data}), provides a consistency check in the bottom row: the M31/M33 placement is not solely an artefact of the CN+TiO 7050\,\angstrom\ blend.  Figure~\ref{fig:locus} fixes only this qualitative two-feature placement; the formal multi-feature $\chi^2$ best-fit grid points of Section~\ref{sec:results-fits} use additional indices (CaT, N\,{\sc i}, and the C$_2$ Swan windows) and may land at different $\Delta\mathrm{C}+\Delta\mathrm{N}$ values when those features pull the joint fit toward grid boundaries.

The LMC grid behaves differently for a physical reason that is itself informative.  At $[\mathrm{Fe/H}]\simeq -0.5$ and $[\alpha/{\rm Fe}]\simeq -0.3$, its absolute oxygen, titanium, and other $\alpha$-element abundances are lower by roughly $0.8\,$dex than at M31 metallicity.  In a C/O$<$1 atmosphere the residual non-CO carbon is the principal feedstock for CN, and a lower O abundance reshapes the C/O ratio and the molecular equilibrium in the direction that flattens the differential response of CN and TiO to $\Delta\mathrm{C}+\Delta\mathrm{N}$ at fixed $T_{\rm eff}$.  Lower [Fe/H] also keeps both species closer to the linear part of the curve of growth, so a $\Delta\mathrm{C}+\Delta\mathrm{N}$ change moves both indices in similar proportion rather than separating them by differential saturation.  Both effects combine to make the iso-$T_{\rm eff}$ curves nearly parallel in (TiO 7050\,\angstrom, CN red) space.  The LMC weak-CN data point still falls on the locus, so the same atmosphere physics applies at lower [Fe/H] and $[\alpha/{\rm Fe}]$, but the optical EW indices are less able to turn the LMC detection into a rotation discriminator.

\subsection{Joint \texorpdfstring{$\chi^2$}{chi-square} Fits and Best-Fit Spectra}\label{sec:results-fits}

The qualitative placement in Figure~\ref{fig:locus} motivates a quantitative joint EW $\chi^2$ over the per-galaxy feature set in Tables~\ref{tab:grid} and \ref{tab:index},
\begin{equation}
\chi^2 =
\sum_k
\left(\frac{{\rm EW}^{\rm model}_k - {\rm EW}^{\rm obs}_k}{\sigma_k}\right)^2 .
\label{eq:chi2}
\end{equation}
For the LMC, the sum includes CN red, TiO 7050\,\angstrom, CaT 8542\,\angstrom, and N~{\sc i} 7442\,\angstrom.  For M31 and M33, where the blue coverage is available, the sum includes CN red, TiO $\gamma$ 6149\,\angstrom, CaT 8542\,\angstrom, N~{\sc i} 7442\,\angstrom, and the two C$_2$ Swan windows.  TiO 7050\,\angstrom\ enters the M31 and M33 locus diagram of Figure~\ref{fig:locus} as a diagnostic comparison, but TiO $\gamma$ is the cleaner TiO anchor in the formal fit because the 7050 window is a CN+TiO blend.  We have nonetheless tested adding TiO 7050\,\angstrom\ to the M33 and M31 $\chi^2$ as a seventh feature: the inferred $\Delta\mathrm{C}+\Delta\mathrm{N}$ stays unchanged for M33 and shifts by $+0.1\,$dex for M31, while the best-fit $T_{\rm eff}$ moves by at most $100\,$K, so the qualitative result is robust to the choice of whether to include the blended index.  The grid search minimises Equation~\ref{eq:chi2} directly over all converged grid points; the $\chi^2$ and per-feature residuals at the best-fit grid point are reported in Table~\ref{tab:fits}.

\begin{table*}[!t]
\centering
\caption{Joint $\chi^2$ best-fit grid points per galaxy.  Columns are the grid parameters at the $\chi^2$ minimum; $\chi^2$ over the per-galaxy feature set (Table~\ref{tab:index}) and number of features used; and the single worst per-feature residual in units of the adopted $\sigma_{\rm EW}$.}
\label{tab:fits}
\small
\begin{tabular}{lrrrrrl}
\toprule
Galaxy & $T_{\rm eff}$ (K) & $[\alpha/{\rm Fe}]$ & $\Delta\mathrm{[C/H]}$ & $\Delta\mathrm{[N/H]}$ & $\chi^2 / N_{\rm feat}$ & Worst-fit feature (residual) \\
\midrule
LMC & 4600 & $-0.3$ & $+0.3$ & $\phantom{+}0.0$ & $1.20/4$ & N~{\sc i}\,7442\,\angstrom\ ($+0.80\sigma$) \\
M33 & 3800 & $\phantom{-}0.0$ & $\phantom{+}0.0$ & $+0.8$ & $2.83/6$ & C$_2$ Swan 5165\,\angstrom\ ($+1.32\sigma$) \\
M31$^{\dagger}$ & 4100 & $-0.3$ & $-0.4$ & $\phantom{+}0.0$ & $1.29/6$ & C$_2$ Swan 5636\,\angstrom\ ($-1.02\sigma$) \\
\bottomrule
\end{tabular}
\tablenotetext{\dagger}{The M31 offset is anchored by the CaT 8542\,\angstrom\ line: dropping CaT from the joint fit moves $\Delta\mathrm{C}+\Delta\mathrm{N}$ from $-0.4$ to $+0.3\,$dex, with the best-fit temperature rising to $T_{\rm eff}=4500\,$K and $[\alpha/{\rm Fe}]=0.0$ (both consistent with the literature M31 disc); see Section~\ref{sec:disc-rotation}.}
\end{table*}

\begin{figure*}[!t]
\centering
\includegraphics[width=\textwidth]{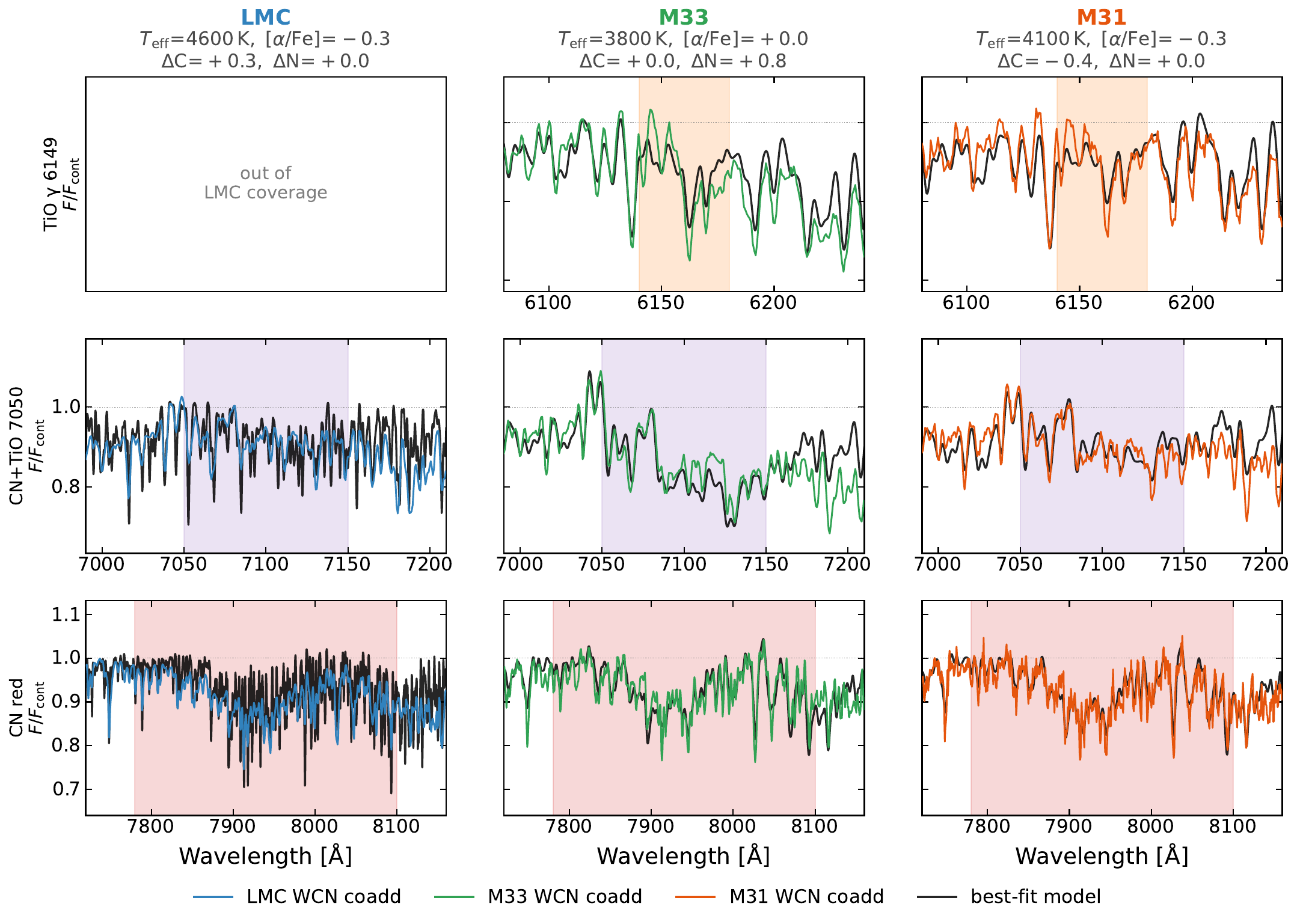}
\caption{Observed weak-CN coadds (coloured) compared to the joint best-fit synthetic spectra (black) at the grid points of Table~\ref{tab:fits}, in the three primary diagnostic windows.  Each model is processed through the same resolution-matched pseudo-continuum recipe as the observation.  The principal result is that the CN, CN+TiO 7050\,\angstrom, and TiO\,$\gamma$ morphologies are reproduced at the adopted precision in all three galaxies by ordinary cool-RSG models of the self-consistent grid, without invoking a carbon-star-like abundance pattern.  The LMC TiO\,$\gamma$ panel is blank because TiO\,$\gamma$\,6149\,\angstrom\ lies outside the $6850$--$9150\,$\angstrom\ CTIO/Hydra coverage of the LMC weak-CN spectra.}
\label{fig:bestfit}
\end{figure*}

Figure~\ref{fig:bestfit} shows the same result in continuum-normalised flux space.  The LMC fit uses the HYDRA-resolution red spectrum, while the M31 and M33 fits use the C$_2$-extended spectra where blue coverage is needed.  The matches are not perfect line-by-line, as expected for low- and medium-resolution molecular spectra, but the observed weak-CN morphology and the fitted diagnostic windows are reproduced at the adopted precision.

\begin{table*}[!htbp]
\centering
\caption{Per-feature residuals at the joint $\chi^2$ best-fit grid points of Table~\ref{tab:fits}, in units of the per-(galaxy, feature) $\sigma_{\rm EW}$ floors of Table~\ref{tab:index}.  Positive entries indicate that the model overpredicts the observed EW (Equation~\ref{eq:chi2}).  Blank entries are not in the fitted feature set for that galaxy (Table~\ref{tab:grid}); the LMC uses TiO\,7050\,\angstrom\ in the fit while M33 and M31 use TiO\,$\gamma$\,6149\,\angstrom\ as the cleaner TiO anchor (Section~\ref{sec:results-fits}).}
\label{tab:residuals}
\small
\begin{tabular}{lrrrrrrr}
\toprule
Galaxy & C$_2$ 5165\,\AA & C$_2$ 5636\,\AA & TiO\,$\gamma$\,6149\,\AA & TiO\,7050\,\AA & N~{\sc i}\,7442\,\AA & CN red & CaT\,8542\,\AA \\
       & \multicolumn{7}{c}{Residual $(\textrm{model}-\textrm{obs}) / \sigma_{\rm EW}$} \\
\midrule
LMC & \nodata & \nodata & \nodata & $-0.05\,\sigma$ & $+0.80\,\sigma$ & $-0.72\,\sigma$ & $-0.19\,\sigma$ \\
M33 & $+1.32\,\sigma$ & $-0.92\,\sigma$ & $-0.49\,\sigma$ & \nodata & $+0.01\,\sigma$ & $-0.08\,\sigma$ & $-0.10\,\sigma$ \\
M31 & $-0.01\,\sigma$ & $-1.02\,\sigma$ & $+0.24\,\sigma$ & \nodata & $+0.12\,\sigma$ & $-0.38\,\sigma$ & $+0.21\,\sigma$ \\
\bottomrule
\end{tabular}
\end{table*}

Three results follow directly from Table~\ref{tab:fits} and Figure~\ref{fig:bestfit}.  The absolute $\chi^2$ values in column 6 of Table~\ref{tab:fits} depend on the adopted $\sigma_{\rm LL}$ (Section~\ref{sec:data}) and should therefore be read as a relative ranking summary across grid points rather than as an absolute goodness-of-fit metric.  The absolute goodness-of-fit is set instead by the per-feature $\sigma$ residuals at the best-fit grid points, listed in Table~\ref{tab:residuals} in units of the adopted $\sigma_{\rm EW}$ floor.  First, the best-fit grid points lie well within the cool-RSG portion of the grid for all three galaxies: $T_{\rm eff}$ from $3800$ to $4600\,$K, $[\alpha/{\rm Fe}]$ matching the literature value of each host to the grid resolution, and surface CNO offsets at the level of the first-dredge-up prediction (Section~\ref{sec:discussion}).  No fitted feature is discrepant from the best model by more than $\sim 1.3\sigma$ in the adopted uncertainty floor (Table~\ref{tab:residuals}); the largest residual is C$_2$ Swan 5165\,\angstrom\ in M33 at $+1.32\sigma$.  In particular, the C$_2$ Swan residuals in M31 and M33, which are the windows that would most directly distinguish a carbon-rich abundance pattern, are all within $\sim 1.3\sigma$ of zero, consistent with C/O$<$1 atmospheres at the inferred surface composition.

Second, the per-galaxy best-fit $[\alpha/{\rm Fe}]$ recovered by the joint EW fit is consistent, at the grid spacing of $0.3\,$dex, with the literature stellar-population values of each host.  The LMC fit returns $[\alpha/{\rm Fe}]=-0.3$, consistent with the sub-solar $\alpha$-enhancement measured for LMC RSGs \citep{dav15}.  The M33 fit returns $[\alpha/{\rm Fe}]=0.0$, consistent with the near-solar disk $\alpha/{\rm Fe}$ inferred from M33 PNe and H\,{\sc ii} regions by \citet{mar09}.  The M31 fit returns $[\alpha/{\rm Fe}]=-0.3$ at the grid minimum, with $\Delta\chi^2\simeq 2$ at the $[\alpha/{\rm Fe}]=0$ slice, so $[\alpha/{\rm Fe}]$ is constrained at roughly the $1.5\sigma$ level and is consistent with the near-solar M31 disk value within that uncertainty.

Third, the surface CNO is constrained along the C+N degeneracy ridge in $(\Delta\mathrm{C},\Delta\mathrm{N})$ rather than as independent offsets.  Reading the individual best-fit grid points literally would suggest $(\Delta\mathrm{C},\Delta\mathrm{N})\simeq(+0.3,0.0)$ for the LMC, $(0.0,+0.8)$ for M33, and $(-0.4,0.0)$ for M31, but as Section~\ref{sec:physics} showed, the optical CN bands respond to the joint C+N reservoir, so the $\chi^2$ surface in $(\Delta\mathrm{C},\Delta\mathrm{N})$ is elongated along the ridge $\Delta\mathrm{C}+\Delta\mathrm{N}=\text{const}$ and the individual values are not separately measurable.

The robust observable is therefore the combined offset $\Delta\mathrm{C}+\Delta\mathrm{N}$, which is $+0.30$, $+0.80$, and $-0.40\,$dex for the LMC, M33, and M31 best-fit grid points respectively.  Extending the grid to $\Delta\mathrm{[C/H]}\in[-0.6,+0.3]$ and $\Delta\mathrm{[N/H]}\in[0.0,+0.8]$ confirms that the ridge is shallow and broad: the formal argmin wanders by 1--2 grid points along the ridge depending on EW noise; $\Delta\chi^2$ along the ridge stays $\lesssim 0.9$ across the full extended span; M33 hits the high-$\Delta\mathrm{N}$ boundary while LMC sits at high $\Delta\mathrm{C}$ with $\Delta\mathrm{N}=0$; and in all three galaxies $\Delta\chi^2(\Delta\mathrm{N}=+0.5)\leq 0.5$.

\begin{figure*}[t]
\centering
\includegraphics[width=\textwidth]{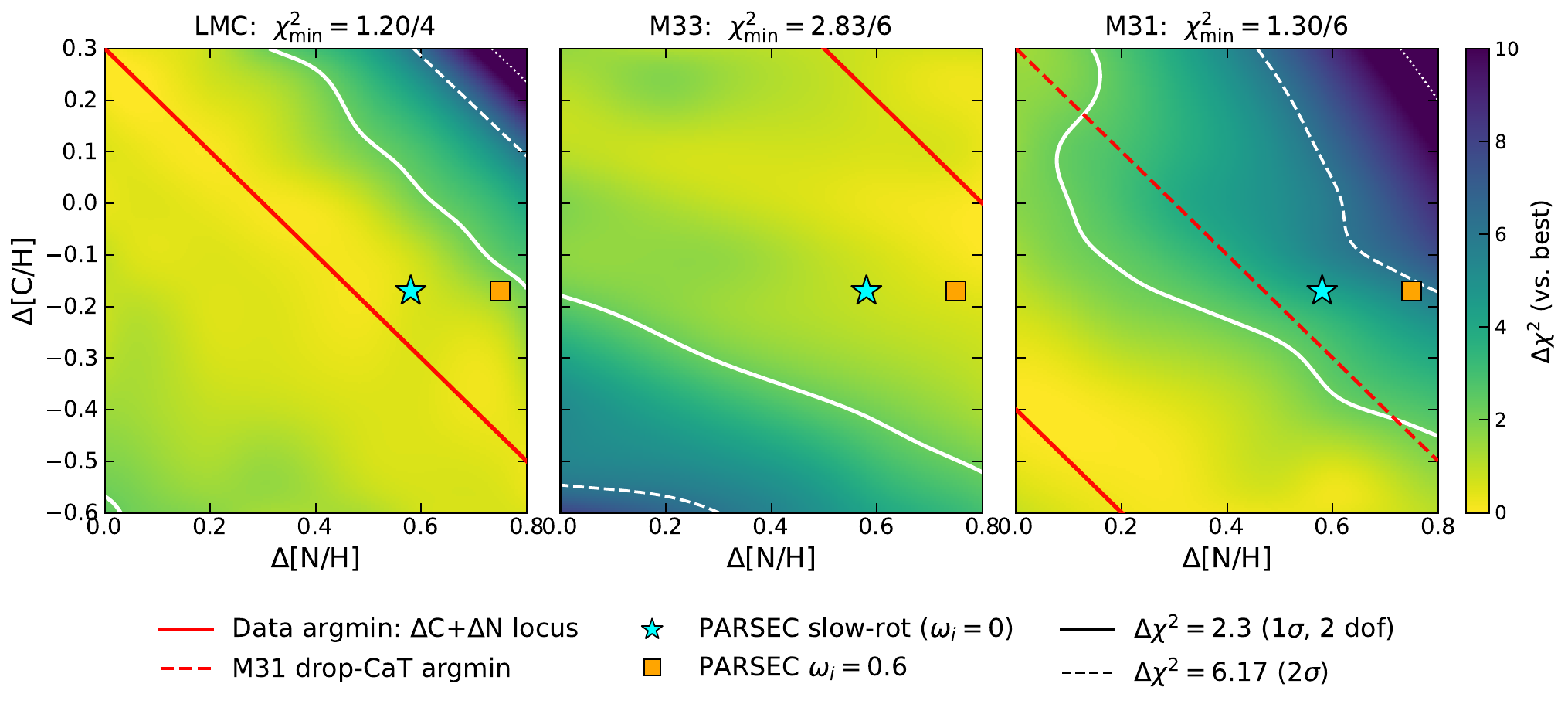}
\caption{Profile $\chi^2$ in the $(\Delta\mathrm{[C/H]}, \Delta\mathrm{[N/H]})$ plane, marginalised over $T_{\rm eff}$ and $[\alpha/{\rm Fe}]$, on the extended grid.  White contours: $\Delta\chi^2=2.3$ (1$\sigma$, 2 dof; solid) and $6.17$ (2$\sigma$; dashed).  PARSEC first-dredge-up predictions are shown as a cyan star ($\omega_i=0$, $\Delta\mathrm{C}=-0.17$, $\Delta\mathrm{N}=+0.58$) and an orange square ($\omega_i=0.6$, $\Delta\mathrm{N}=+0.75$).  Because the data constrain only the combined offset $\Delta\mathrm{C}+\Delta\mathrm{N}$, we plot the data argmin as a solid red iso-$(\Delta\mathrm{C}+\Delta\mathrm{N})$ locus rather than a point.  For LMC and M33 the locus and the PARSEC slow-rotation point lie within $1\sigma$ of each other.  For the baseline M31 fit the locus sits at $\Delta\chi^2\simeq 3.8$ ($\sim 2\sigma$) below PARSEC; the dashed red locus is the M31 argmin after CaT is dropped from the fit (Section~\ref{sec:results-ablations} (iv)) and crosses the PARSEC point inside $1\sigma$.}
\label{fig:chi2ridge}
\end{figure*}

Figure~\ref{fig:chi2ridge} shows the $\chi^2$ surface in $(\Delta\mathrm{C}, \Delta\mathrm{N})$ directly.  Because the data constrain only $\Delta\mathrm{C}+\Delta\mathrm{N}$, we draw the data constraint as a solid red iso-$(\Delta\mathrm{C}+\Delta\mathrm{N})$ locus through the argmin rather than a single point.  For LMC and M33 the red and cyan loci (the latter passing through the PARSEC slow-rotation point) are close and both lie inside the $1\sigma$ contour.  For M31 the baseline solid red locus sits below the cyan one; the dashed red locus shows where the argmin moves when CaT is dropped from the joint fit (Section~\ref{sec:results-ablations} (iv)) and overlaps the cyan slow-rotation locus.  The implications for rotation are taken up in Section~\ref{sec:disc-rotation}.

\subsection{Robustness Ablations}\label{sec:results-ablations}

We test the robustness of the joint EW $\chi^2$ formulation through five orthogonal ablations of feature choice, integration scale, and synthesis assumptions.

(i) \emph{Sub-feature decomposition.}  We split the CN red 7780--8100\,\angstrom\ window into four 80\,\angstrom\ sub-bands and recompute per-sub-band EWs, to test whether the integrated CN red EW is absorbing a within-window gradient.  The rms residual across the four sub-bands at the best-fit grid point is $\sim 1.4\,\sigma$ for LMC, $\sim 0.9\,\sigma$ for M33, and $\sim 0.5\,\sigma$ for M31, and the per-sub-band residuals share sign across the four sub-windows in each galaxy.  The integrated CN red EW is therefore tracking an approximately uniform CN scaling rather than a hidden sub-band gradient.

(ii) \emph{Pixel-level residual.}  At the best-fit grid point, the continuum-normalised pixel rms residual is $\sim 0.02$--$0.10$ across the diagnostic windows in all three galaxies --- the order-of-magnitude precision expected for molecular spectra at these resolutions.  The corresponding integrated EW residuals, listed in Table~\ref{tab:residuals}, stay within $\pm 1.3\,\sigma_{\rm EW}$ for every fitted feature in all three galaxies.  The contrast --- non-trivial per-pixel residual but small bulk-EW residual --- is the empirical signature that the line-list and pseudo-continuum-recipe noise that dominates the pixel-level mismatch (Section~\ref{sec:data}) averages out under integration into the EW, which is the formal motivation for the EW $\chi^2$ formulation made precise in ablation~(iii) below.

(iii) \emph{Pixel residual vs.\ parameter shift.}  We compare the pixel-level and EW-level responses to a one-grid-step shift in $\Delta\mathrm{N}$.  At the pixel level, the per-window pixel rms changes by less than $20\%$ for LMC and less than $5\%$ for M33 and M31 --- the per-pixel residual is dominated by line-list and pseudo-continuum-recipe noise (Section~\ref{sec:data}) rather than by the parameter mismatch.  At the EW level, by contrast, integration over the feature window suppresses the uncorrelated per-pixel noise while preserving the coherent bulk shift induced by the parameter change, and the resulting EW shift is large enough relative to $\sigma_{\rm EW}$ to give the EW $\chi^2$ a sharp, well-defined minimum in the direction perpendicular to the C$+$N degeneracy ridge.  The EW $\chi^2$ therefore discriminates parameters cleanly even though a pixel $\chi^2$ would not, which is why we adopt the EW formulation as the fitting statistic.

(iv) \emph{Feature drop.}  We refit after removing each feature in turn.  For LMC the inferred $\Delta\mathrm{C}+\Delta\mathrm{N}$ stays within $\pm 0.2\,$dex of the full-fit value across all drops.  Two single-feature drops matter for M33 and M31.  The baseline M33 fit lands at the upper $\Delta\mathrm{N}$ grid edge ($\Delta\mathrm{N}=+0.8\,$dex; Table~\ref{tab:fits}); dropping TiO\,$\gamma$\,6149\,\angstrom\ pulls $\Delta\mathrm{C}+\Delta\mathrm{N}$ down from $+0.8$ to $+0.3\,$dex, close to the PARSEC slow-rotation value.  The baseline M31 fit sits at low $\Delta\mathrm{C}+\Delta\mathrm{N}=-0.4\,$dex; dropping CaT\,8542\,\angstrom\ moves $\Delta\mathrm{C}+\Delta\mathrm{N}$ from $-0.4$ to $+0.3\,$dex, matching the PARSEC slow-rotation value.  These two single-feature ablations therefore identify the individual features responsible for the M33 grid-edge result and the M31 tension with PARSEC, and we take them up in Section~\ref{sec:disc-rotation}.

(v) \emph{Microturbulence.}  Motivated by the LMC RSG measurements of \citet{dav15} giving $\xi\simeq 3$--$4\,$km\,s$^{-1}$ relative to the $\xi=2\,$km\,s$^{-1}$ adopted here, we re-synthesised the full LMC grid at $\xi=3.5\,$km\,s$^{-1}$.  The LMC CN red, TiO 7050\,\angstrom, Ca\,{\sc ii} triplet, and N~{\sc i} EWs at $R=7000$ shift by less than $0.05\,$\angstrom\ in the vast majority of grid points across the $(T_{\rm eff},\Delta\mathrm{C},\Delta\mathrm{N})$ plane, and the formal argmin and the $\chi^2$ at the PARSEC slow-rotation locus are unchanged at the $\Delta\chi^2\lesssim 0.1$ level.  The physical reason is that at $R\simeq 7000$ the CN red EW is set by the number of unsaturated rotational lines integrated across the band rather than by atomic-line saturation cores, so microturbulence broadening barely shifts the molecular bandhead EW used here.

The dominant remaining uncertainty on $\Delta\mathrm{C}+\Delta\mathrm{N}$ in any single galaxy is therefore the EW systematic floor (recipe plus line-list, Section~\ref{sec:data}), not the photon noise (which the inverse-variance budget of the coadds places at $\lesssim 0.1\,$\angstrom\ per feature, well below the $\geq 1\,$\angstrom\ floor) and not the microturbulence choice.  The best-fit temperatures should accordingly be read as grid-level summaries of coadded populations at fixed $\log g$ and microturbulence, not as precision effective temperatures for individual stars, at the $\pm 100$--$200\,$K level set by the grid spacing and the same systematic floors.

\section{Discussion}\label{sec:discussion}

\subsection{Weak-CN Stars Are Ordinary Cool RSGs}\label{sec:disc-puzzle}

The per-galaxy fits of Section~\ref{sec:results} place all three weak-CN coadds inside the ordinary cool-RSG region of the self-consistent grid, with a host-dependent rotation discrimination.  The discovery papers compared weak-CN stars to carbon stars because the broad CN band was the selection clue.  In classical optical classification, strong CN and C$_2$ Swan are associated with C-rich atmospheres, whereas TiO and the Ca\,{\sc ii} triplet are associated with non-carbon, C/O$<$1 cool stars.  Weak-CN stars combine these signatures: weak carbon-star-like CN, but TiO/CaT and young-RSG photometric positions.

The apparent contradiction is resolved once the CN band is treated as a normal molecular feature of a C/O$<$1 cool-RSG atmosphere rather than as a carbon-star flag.  With the self-consistent line-blanketing back-reaction in place --- the iterated grid of Section~\ref{sec:grid} rather than a pre-computed scaled-solar grid \citep{ari16} --- the mixed appearance is the expected spectrum of a cool RSG in the relevant temperature regime, and the same physical reasoning that goes back to \citet{sca74}'s band-strength analysis across the M, MS, and S spectral sequence --- the chain of cool giants whose dominant metal-oxide bands shift smoothly from TiO toward ZrO as the C/O ratio approaches unity --- applies in the cool-RSG regime as well.

An independent line of evidence pointing to the same conclusion appears in the Comparison to Empirical Templates (COMET) classification framework of \citet{guh25} and \citet{gri25}.  COMET scores each spectrum by its modified-$\chi^2$ similarity against two empirical templates --- a weak-CN coadd and a carbon-star coadd --- evaluated over the $7800$--$8200\,$\angstrom\ wavelength range.  In the resulting $(\chi^2_{\rm wCN}, \chi^2_{\rm C})$ plane, the visually-identified carbon stars form a distinct group in the lower-right, while the remainder of the sample traces a comet-like distribution with the weak-CN coadds at the head and normal O-rich stars spreading along the tail.  The COMET geometry therefore places the weak-CN class at the high-CN tip of the O-rich (C/O$<$1) distribution rather than as an intermediate between normal and carbon stars.  Our atmospheric-modelling result --- that weak CN is the warm-end molecular-equilibrium signature of an ordinary C/O$<$1 cool RSG --- is consistent with this COMET picture: the weak-CN class is the empirical handle on the bright end of the normal cool-RSG sequence.

\subsection{Slow-Rotation Consistency and the Formation Channel}\label{sec:disc-rotation}

For reference, the PARSEC v2.0 \citep{cos19a,ngu22} surface mass-fraction columns at the core-He-burning phase for $5$--$10\,\mathrm{M}_\odot$ progenitors at $Z=0.014$ give first-dredge-up offsets $\Delta\mathrm{[C/H]}\simeq -0.17$, $\Delta\mathrm{[N/H]}\simeq +0.58$ at $\omega_i=0$, $\Delta\mathrm{[N/H]}\simeq +0.75$ at $\omega_i=0.6$, and $\Delta\mathrm{[N/H]}\simeq +0.85$ to $+0.95$ at $\omega_i=0.9$.  The corresponding $\Delta\mathrm{C}+\Delta\mathrm{N}$ moves only weakly with $\omega_i$ because rotational mixing simultaneously raises N and depresses C: $+0.41$, $+0.48$, $+0.45$ at $\omega_i=0,0.6,0.9$ respectively.  This is also the physical content of the thermohaline-plus-rotation models of \citet{cha10} and \citet{lag12} at lower progenitor masses, extended here to the cool-RSG regime: surface CNO is a product of dredge-up plus rotational mixing acting together, and the optical CN band integrates them.

Mapping our measured $\Delta\mathrm{C}+\Delta\mathrm{N}$ values onto initial rotation rate carries an intrinsic limitation: under PARSEC v2.0, $\Delta\mathrm{C}+\Delta\mathrm{N}$ changes by only $\simeq 0.07\,$dex between $\omega_i=0$ and $\omega_i=0.6$, because rotational mixing simultaneously raises N and depresses C.  The variable we measure robustly is therefore weakly rotation-dependent by construction, and the EW fit alone cannot cleanly discriminate slow- from fast-rotation channels.  What the fit can say is that the data are consistent with slow-rotation first dredge-up in LMC and M33 ($\Delta\chi^2(\omega_i=0)\lesssim 0.9$; Figure~\ref{fig:chi2ridge}), while the baseline M31 fit sits at $\sim 2\sigma$ below the slow-rotation locus along the $\Delta\mathrm{C}+\Delta\mathrm{N}$ axis.

The M31 offset is anchored by the CaT 8542\,\angstrom\ line: removing CaT from the joint fit shifts the M31 argmin onto the iso-$(\Delta\mathrm{C}+\Delta\mathrm{N})$ locus that runs through the PARSEC slow-rotation point, inside the $1\sigma$ contour (Figure~\ref{fig:chi2ridge}, dashed red locus in the M31 panel), so the formal $2\sigma$ tension is a single-feature calibration question rather than a CNO physics one.

The rotation interpretation is therefore best read as a conditional inference from the EW fits plus the adopted stellar-evolution prior, not as a direct measurement of $v\sin i$ or of an underlying initial-rotation distribution.  Standard single-star evolution with first dredge-up reproduces the CNO signature observed here in LMC and M33 without invoking binary spin-up channels, and the data do not require the larger $\Delta\mathrm{[N/H]}$ that such channels would generically produce.

The independent constraint that selects against fast rotators in this sample comes from the visual ``weak-CN'' cut itself, not from the EW fit: the young Magellanic Cloud cluster NGC 1866 hosts a red main-sequence population of fast rotators near critical rotation ($\omega\simeq 0.9\,\omega_c$) that contains the majority of the cluster's main-sequence stars at the relevant progenitor masses \citep{milone17}, and the broader thermohaline-plus-rotation surface-CNO literature \citep{cha10,lag12,cos19a} predicts that fast rotators produce stronger CN once dredge-up is complete.  A low-CN-EW cut by construction excludes the strong-CN tail of the rotation distribution, so the weak-CN coadd is biased toward the slow-rotation end of the underlying RSG population.  This sample-construction argument is independent of, and complementary to, the C+N EW measurement: the EW fit measures $\Delta\mathrm{C}+\Delta\mathrm{N}$, while the visual cut filters the rotation distribution at the input.

Several additional systematics could further shift the M31 baseline result beyond the CaT calibration identified above: an M31 sample skewed toward lower-mass or pre-first-dredge-up progenitors, a fractional underestimate of the 7400--8400\,\angstrom\ pseudo-continuum at solar metallicity (where the metallic line blanketing is deepest), or a $\pm 0.1\,$dex bias in the adopted M31 O abundance that shifts the C+N degeneracy ridge in the $\Delta\mathrm{C}+\Delta\mathrm{N}$ direction.  Discriminating these requires higher-resolution M31 spectroscopy to break the C+N degeneracy directly, per-star photometric ages or masses for the M31 weak-CN sample, an independent calibration of the M31 pseudo-continuum recipe against stars of known surface composition, or a re-validation of the CaT line strength against Galactic-disc RSGs of known $T_{\rm eff}$ and $[\alpha/{\rm Fe}]$.

\subsection{Environment, Incidence, and Recent Star Formation}\label{sec:disc-sfr}

The atmospheric-modelling result of Section~\ref{sec:results} --- that weak CN is the molecular-equilibrium signature of an ordinary He-burning cool RSG of $5$--$10\,\mathrm{M}_\odot$ --- enables one corollary application that does not depend on resolving $\Delta\mathrm{C}$ and $\Delta\mathrm{N}$ separately: weak CN remains a useful tracer even where the optical C+N abundance constraint is degeneracy-limited.

The discovery papers place the sample on the cool-RSG core-He-burning locus at masses of $5$--$10\,\mathrm{M}_\odot$ \citep{guh25,gri25}, so the population-level incidence should track the recent ($\lesssim 40\,$Myr-averaged) star formation rate on the timescale set by the He-burning lifetime of those progenitors; weak-CN counts per unit host stellar mass are equivalently a specific star formation rate (sSFR) tracer over the same timescale.  A simple stellar-evolution scaling relates the expected weak-CN count to the recent SFR through
\begin{equation}
N_{\rm WCN} \;\propto\;
\langle\mathrm{SFR}\rangle_{40\,{\rm Myr}} \;
\tau_{\rm RSG} \;
\frac{f_{\rm IMF}(5\text{--}10\,\mathrm{M}_\odot)}{\langle m\rangle} \, ,
\label{eq:sfr}
\end{equation}
where $\langle\mathrm{SFR}\rangle_{40\,{\rm Myr}}$ is the SFR averaged over the $\sim 40\,$Myr main-sequence lifetime of the progenitor mass range, $\tau_{\rm RSG}$ is the duration of the cool He-burning plateau, $f_{\rm IMF}(5\text{--}10\,\mathrm{M}_\odot)$ is the IMF mass fraction in the progenitor range (a few per cent for standard IMFs), and $\langle m\rangle$ is the mean progenitor mass.  Ratios of weak-CN counts across hosts should therefore approximate ratios of their recent SFRs, modulated by spectroscopic survey selection.

Published recent SFRs of the three hosts span a factor of two: the LMC at $\approx 0.2\,\mathrm{M}_\odot\,{\rm yr}^{-1}$ \citep{harris09,zar04}, M33 at $\sim 0.3\,\mathrm{M}_\odot\,{\rm yr}^{-1}$ \citep{will21,lazz22}, and M31 at $\sim 0.4\,\mathrm{M}_\odot\,{\rm yr}^{-1}$ \citep{lewi15}.  The survey-level weak-CN counts are $N_{\rm LMC}=779$, $N_{\rm M33}=659$, $N_{\rm M31}=224$ \citep{guh25,gri25}.  A naive application of Equation~\ref{eq:sfr} with the LMC value used to absorb the common $\tau_{\rm RSG}f_{\rm IMF}/\langle m\rangle$ prefactor predicts $N^{\rm pred}_{\rm M33} = 779 \times (0.3/0.2) \simeq 1170$ and $N^{\rm pred}_{\rm M31} = 779 \times (0.4/0.2) \simeq 1560$, both substantially larger than the observed counts.  Equivalently, if one inverts Equation~\ref{eq:sfr} to read recent SFR off the observed weak-CN counts at fixed LMC anchor, the inferred values are $\sim 0.17\,\mathrm{M}_\odot\,{\rm yr}^{-1}$ for M33 and $\sim 0.06\,\mathrm{M}_\odot\,{\rm yr}^{-1}$ for M31 --- under by factors of $\sim 2$ and $\sim 7$ relative to the literature SFRs respectively.

The discrepancy direction (counts under-predicted at larger host distance) is what one expects from spectroscopic survey selection rather than from a breakdown of the molecular-physics picture.  The LMC HYDRA programme observed a wide-field fibre target list, while the M31/M33 DEIMOS programme drew from a much smaller mask footprint at much greater distance and with a more aggressive target-priority weighting against the cool-RSG colour locus.  A quantitative test of Equation~\ref{eq:sfr} would have to forward-model the survey selection per host, which is beyond the present scope.  At order-of-magnitude level the weak-CN counts and the host SFRs are nevertheless consistent with the population being a $T_{\rm eff}$-selected sub-class of ordinary He-burning RSGs rather than a chemically peculiar tracer requiring its own formation channel.

\subsection{The Missing Pure-Carbon Lever and Outstanding Questions}\label{sec:disc-c2}

The single outstanding observational lever in this analysis is a pure-carbon indicator that could anchor $\Delta\mathrm{C}$ separately from $\Delta\mathrm{C}+\Delta\mathrm{N}$ and so break the C+N degeneracy.  The C$_2$ Swan windows (5125--5180 and 5585--5640\,\angstrom) are the obvious candidates: they are present in the M31/M33 grid and are measured for both galaxies; the residuals in Table~\ref{tab:fits} are within $\sim 1.3\sigma$, consistent with C/O$<$1 atmospheres at the inferred surface composition, but the $\sigma_{\rm C_2}\simeq 2.5$--$3.4\,$\angstrom\ floor at DEIMOS $R\simeq 2000$ is wide, and is dominated by the 20\% line-list calibration term rather than by recipe drift.  Reducing it therefore requires a model-side improvement in the C$_2$ Swan line list combined with a per-window continuum calibration, or a higher-resolution measurement in which the broad molecular bands become resolved against the pseudo-continuum rather than smearing into it.

A further caveat that the present analysis cannot internally test is the assumption of plane-parallel atmospheric geometry.  \texttt{pykurucz} uses the ATLAS-style plane-parallel formulation, while at $\log g\simeq 0$ the photospheric extension is a non-negligible fraction of the stellar radius, and spherical-symmetric models more accurately capture the radiative transfer through the upper layers.  The leading-order effect of sphericity on the integrated band EWs used here at $R\leq 7000$ is expected to fall within the adopted $\sigma_{\rm EW}$ floors for most indices, but we cannot test this internally.  A definitive treatment will require re-fitting with spherical-symmetric atmospheres (e.g., COMARCS; \citealt{ari16}), which we identify as the natural next step beyond the present analysis.

A direct test of the rotation channel is the surface $^{12}$C/$^{13}$C ratio in the CN bandheads.  Rotation-driven mixing brings up $^{13}$C-enriched material from the CNO-burning shell along with raising the absolute N abundance, and the resulting $^{13}$CN bandheads imprint a wavelength-resolved pattern on the CN red region rather than a bulk EW shift.  We have synthesised this prediction at the LMC anchor by varying the $^{12}$CN/$^{13}$CN partition in the \citet{brooke14} CN red-system line list while holding the rest of the atmosphere fixed: the maximum flux deviation between the solar-isotope reference and the strongly mixed case ($^{12}$C/$^{13}$C $\simeq 5$) reaches $\sim 1.8\%$ near $7950\,$\angstrom\ and falls below $0.5\%$ outside the narrow $7900$--$8100\,$\angstrom\ window where the $^{13}$CN bandheads cluster.  At the spectral resolutions of the present data ($R\simeq 2000$--$7000$) this is unresolvable; at $R\gtrsim 30\,000$ on bright Galactic RSG analogues it is the natural next-step measurement.

\section{Conclusions}\label{sec:conclusion}

We have reanalyzed the weak-CN spectra of \citet{guh25} for M31 and M33 and \citet{gri25} for the LMC using a four-dimensional, self-consistently iterated cool-RSG model atmosphere grid in \texttt{pykurucz} \citep{kim26} spanning $T_{\rm eff}$, $[\alpha/{\rm Fe}]$, $\Delta\mathrm{[C/H]}$ and $\Delta\mathrm{[N/H]}$ at each host's metallicity.  Our main conclusions are:

\begin{enumerate}
\item Weak CN is not evidence for a carbon-star intermediate phase.  It forms naturally in ordinary C/O$<1$ cool-RSG atmospheres near the CN/TiO molecular transition --- TiO crosses its dissociation transition near $T_{\rm eff}\sim 3800\,$K, while CN remains present up to $T_{\rm eff}\sim 4300\,$K, so a shallow CN band in an otherwise non-carbon-star spectrum is the expected morphology for a cool RSG inside this transition window.
\item Because the optical CN bands respond to the product of available C and N number densities, the data constrain the combined C+N surface abundance $\Delta\mathrm{[C/H]}+\Delta\mathrm{[N/H]}$ rather than each abundance individually.  The C$_2$ Swan bands were tested as a candidate pure-carbon lever; at DEIMOS $R\simeq 2000$ they do not yet split the C/N degeneracy at the line-list-dominated systematic floor.
\item Joint $\chi^2$ best-fit grid points reproduce the observed weak-CN coadds with ordinary RSG parameters and per-feature residuals at the level of the adopted EW systematic floors.  Across the three galaxies the combined C+N surface abundance $\Delta\mathrm{[C/H]}+\Delta\mathrm{[N/H]}$ is $+0.3$, $+0.8$, and $-0.4\,$dex for the LMC, M33, and M31 best-fit grid points respectively (or $+0.3\,$dex for M31 when CaT is dropped from the fit; see point 4 below); the PARSEC slow-rotation value ($\Delta\mathrm{C}+\Delta\mathrm{N}\simeq +0.4$) sits within $\Delta\chi^2 \lesssim 0.9$ of all three argmins under that reading.
\item Under the PARSEC v2.0 \citep{cos19a,ngu22} first-dredge-up plus rotational-mixing prior, all three weak-CN coadds are consistent with slow-rotation first dredge-up, and the visual weak-CN cut at the input independently selects against fast rotators.  We cannot, however, rule out faster rotation channels from the EW fit alone, because $\Delta\mathrm{[C/H]}+\Delta\mathrm{[N/H]}$ shifts by only $\simeq 0.07\,$dex between $\omega_i=0$ and $\omega_i=0.6$ under PARSEC.
\end{enumerate}

Weak-CN stars are ordinary cool RSGs; the diagnostic is a temperature window, not a chemical peculiarity.  The TiO--CN dissociation offset and the iterated atmosphere grid together resolve the discovery puzzle without invoking carbon-star chemistry, and place the three weak-CN coadds in the molecular regime where slow-rotation first dredge-up reproduces the observed combined CNO surface abundance.  What weak CN inherits from this picture is well-defined mass and age: a $\sim 5$--$10\,\mathrm{M}_\odot$ cool He-burning sub-class of the RSG population, identified at low spectral resolution by photospheric thermodynamics rather than by surface composition.

\section*{Acknowledgements}

Large-language-model coding assistants (Anthropic Claude Opus 4.7 and OpenAI GPT-5.5) were used during this work to refactor analysis scripts, iterate on prose, and --- the use we found most valuable --- canvas variant assumptions on the EW recipe, $\sigma$ floors, feature choice, and microturbulence at a speed that allowed us to confirm the qualitative conclusions are robust to those choices.  The model output also contained substantive errors and dead ends; sifting them, identifying the load-bearing physical explanations, and constructing the argument of the paper were human tasks, and the scientific results and their interpretation are the authors'.

\bibliography{biblio.bib}

\end{document}